\documentclass[aps,showpacs,showkeys,reprint]{revtex4-1}

\usepackage{amsmath,amssymb,bm}
\usepackage{graphicx}
\usepackage{epstopdf}
\usepackage{dcolumn}
\usepackage{color}
\usepackage{natbib}

\begin{document}

\allowdisplaybreaks

\title{On the use of aerogel as a soft acoustic metamaterial for airborne sound}
%%%%%%
\author{Matthew D. Guild}
\email{matthew.guild.ctr@nrl.navy.mil}
\altaffiliation{Current address: NRC Research Associateship Program, U.S. Naval Research Laboratory, Code 7160, Washington DC 20375, USA}
\affiliation{Wave Phenomena Group, Departamento de Ingenier\'{i}a Electr\'{o}nica,
Universidad Polit\'{e}cnica de Valencia, Camino de vera s/n (Edificio 7F), E-46022 Valencia, Spain}
%%%%%%
\author{Victor M. Garc\'{i}a-Chocano}
\affiliation{Wave Phenomena Group, Departamento de Ingenier\'{i}a Electr\'{o}nica,
Universidad Polit\'{e}cnica de Valencia, Camino de vera s/n (Edificio 7F), E-46022 Valencia, Spain}
%%%%%%
\author{Jos\'{e} S\'{a}nchez-Dehesa}
\email{jsdehesa@upv.es}
\affiliation{Wave Phenomena Group, Departamento de Ingenier\'{i}a Electr\'{o}nica,
Universidad Polit\'{e}cnica de Valencia, Camino de vera s/n (Edificio 7F), E-46022 Valencia, Spain}
%%%%%%
\author{Theodore P. Martin}
\affiliation{U.S. Naval Research Laboratory, Code 7160, Washington DC 20375, USA}
%%%%%%
\author{David C. Calvo}
\affiliation{U.S. Naval Research Laboratory, Code 7160, Washington DC 20375, USA}
%%%%%%
\author{Gregory J. Orris}
\affiliation{U.S. Naval Research Laboratory, Code 7160, Washington DC 20375, USA}
%%%%%%
\date{\today}

\begin{abstract}
Soft acoustic metamaterials utilizing mesoporous structures have recently been proposed as a novel means for tuning the overall effective properties of the metamaterial and providing better coupling to the surrounding air.  In this work, the use of silica aerogel is examined theoretically and experimentally as part of a compact soft acoustic metamaterial structure, which enables a wide range of exotic effective macroscopic properties to be demonstrated, including negative density, density-near-zero and non-resonant broadband slow sound propagation.  Experimental data is obtained on the effective density and sound speed using an air-filled acoustic impedance tube for flexural metamaterial elements, which have previously only been investigated indirectly due to the large contrast in acoustic impedance compared to that of air.  Experimental results are presented for silica aerogel arranged in parallel with either 1 or 2 acoustic ports, and are in very good agreement with the theoretical model.
\end{abstract}

\maketitle

%%%%%%%%%%%%%%%%%%%%%%%%%%%%%%%%%%%%%%%%%
\section{Introduction} \label{Sec:Intro}
Acoustic metamaterials have received interest in recent years by enabling macroscopic physical characteristics which cannot be obtained with traditional materials, such as negative, near-zero or anisotropic dynamic effective fluid properties.  Acoustic metamaterials are able to achieve such previously unattainable exotic properties through the careful design of the microstructure, which create microscale dynamics that result in the desired macroscopic properties. The reader is addressed to the recent reviews on this topic where one can find many of the exciting applications of acoustic metamaterials\cite{maldovan2013review,kadic2013review}.  

Until recently, such acoustic metamaterials have relied on materials which are much harder than the surrounding fluid medium, often treated as acoustically rigid or nearly-rigid structures for airborne sound.  Alternatively, soft acoustic metamaterials utilizing mesoporous structures have been proposed as a novel means for the molding and tuning of the overall properties of the resulting metamaterial, while simultaneously providing better coupling with the acoustic environment around it \cite{Brunet2013}.  
Building upon this concept, the use of silica aerogel as part of a compact and conformal soft acoustic metamaterial structure is examined theoretically and experimentally, yielding an interesting suite of useful yet exotic properties.

In addition to extremely large and/or negative dynamic properties, there are a wide range of interesting and exciting phenomena associated with effective properties that are near zero, particularly those associated with \emph{extraordinary transmission}, which can be achieved when either the effective density or wave speed approaches zero\cite{Christensen2008,Park2013}.  In the case of a density-near-zero material, the effective wave speed increases dramatically and leads to a quasi-static field within a given structure, which can exhibit a supercoupling effect through long narrow channels\cite{Fleury2012,Fleury2013,graciasalgadoPRB2013negative}.  At the opposite extreme, there are also interesting effects which arise as the effective acoustic wave speed approaches zero, which is referred to as \emph{slow sound}, the analogue of \emph{slow light} in optics.  Previous demonstrations have utilized resonant effects using either sonic crystals or detuned resonators\cite{Santillan2011, Robertson2004, Cicek2012}, resulting in slow sound that occurs over a relatively narrow bandwidth.  A novel application of slow sound propagation was recently proposed for the improved design of acoustic absorbers by Groby \emph{et al.}\cite{Groby2015}, in which slow sound in large slits filled with absorptive foam were used to significantly increase the low frequency absorption in air.  

One of the fundamental aspects that gives a metamaterial its exotic macroscopic properties is the homogenization of the microstructure, which has recently been explored for elastic and flexural metamaterial components\cite{Torrent2014, Li2014}.  This is particularly important because the effective macroscopic properties of an acoustic metamaterial can be significantly different than those of the constituent microstructural elements.  When there is open flow through the structure, such as a sonic crystal lattice\cite{Cervera2001,Guild2015} or transmission-line arrangement of Helmholtz resonators \cite{Fang2006}, it is relatively straightforward to extract effective properties experimentally due to the relatively low acoustic impedance.  

A formidable challenge, however, arises when obtaining the effective properties of a metamaterial sample containing elastic elements, which have acoustic impedances that are orders of magnitude greater than the surrounding fluid and at frequencies well below those typically used to obtain acoustic properties through direct time-of-flight measurements.  As a result, previous works in air have either been restricted to theoretical and numerical evaluation\cite{Christensen2008,Bongard2010,Xiao2012,Fleury2013,Oudich2014} or limited to an indirect comparison of the metamaterial properties using experimental results for the reflected and transmitted pressure field\cite{Lee2010,Park2013}.  In this work we make the significant step of experimentally extracting the effective dynamic properties (density and sound speed) of these flexural metamaterial elements, which has to the authors knowledge never been accomplished previously for such a metamaterial structure in air.    

It is expected that the elasticity of the materials defining the metamaterial structure might play a fundamental role in order to understand the phenomena observed in sound transmission and reflectance through the channels defined by the structure. 
In fact, the role of the elastic properties is paramount for the case of structures embedded in water, as it has been recently demonstrated \cite{bozhkoPRB2015redirection}.
Although some recent work has begun to incorporate the elastic effects into the metamaterial structure, many of these designs continue to have the primary dynamic element consisting of mass-spring resonators which are affixed to an elastic plate as structural support\cite{Oudich2014, Xiao2012}.  Alternatively, soft acoustic metamaterials represent a paradigm shift beyond this framework by creating the dynamics from the structure itself.  It is important to emphasize that such a soft acoustic metamaterial, realized with the unique properties of aerogels, can be tailored to obtain a wide spectrum of desirable exotic properties in a single versatile subwavelength acoustic metamaterial element.  
In this work, theoretical and experimental results for a compact metamaterial configuration are presented, enabling a thin, conformal configuration to be realized.  In particular, these structures represent a soft acoustic metamaterial, which are realized using the flexural resonance of the zeroth-order anti-symmetric Lamb-wave mode in silica aerogel disks.  In addition to its sub-wavelength thickness, extreme effective properties are demonstrated across a broad range of the operating bandwidth, with distinct regions exhibiting negative density, density-near-zero, and ultra-low sound speeds.

%%%%%%%%%%%%%%%%%%%%%%%%%%%%%%%%%%%%%%%%%
\section{Background on aerogels} \label{Sec:Background}

While the exceptional thermal properties of aerogel have led to a revolution in thermal insulation applications, utilization of the unique acoustic characteristics have been minimal, primarily relating to marginal improvements of existing concepts such as quarter-wavelength impedance matching or ultrasonic absorbers\cite{Hrubesh1998,AerogelHandbookChp33,Gibiat1995}.  However, aerogels offer several unique features that enable it to function as a subwavelength flexural element in a soft acoustic metamaterial, which are made possible by its unique microstructure.  One of the most common types of aerogels, silica aerogel, consists of a high-porosity frame made of fused silica nanoparticles.  The most notable characteristic of silica aerogel is its extremely low static density which is directly related to the very high porosity of the structure, making it much closer to that of air compared with any other type of elastic solid.  Due to the nanoscale pore size, however, the air is locked in place by viscous effects producing a higher acoustic density than that compared to typical porous media used in acoustic applications\cite{Forest1998}.  Furthermore, the small cross-section connecting the fused nanoparticles results in a very low elastic stiffness, compared with a rigid silica structure of the same porosity\cite{Gronauer1986}.  This combination gives a relatively low acoustic impedance (for an elastic solid), and in particular yields an exceptionally low flexural wave speed, making it ideal for use as a subwavelength flexural element for airborne sound.

When the wavelength is much larger than the microstructure, negative effective properties are achieved via control of the microstructure arrangement and the resulting dynamics.  This feature in the microstructure design is typically achieved with two main types of arrangements: either as a mass-spring system, or in a transmission-line consisting of mass and stiffness elements.  The mass-spring systems, which demonstrate extreme effective mass and stiffness in the vicinity of the mass-spring resonance, are therefore referred to as \emph{locally resonant acoustic metamaterials} (LRAMs) \cite{Liu2000,Naify2010,Naify2011,Ma2013,Park2013,Chen2014}.  Although such mass and spring elements can be arranged in a compact configuration and are relatively simple and robust, the resulting extreme effective properties are inherently narrowband and subject to appreciable loss due to the close proximity of the mass-spring resonance\cite{graciasalgadoPRB2013negative,torrent2011}.   Alternatively, acoustic metamaterials have been proposed using thin elastic plates as a means for operating as a positive stiffness element in acoustic transmission-line arrangements \cite{Bongard2010}, which has recently been applied to acoustic metamaterial leaky-wave antennas\cite{Naify2013}.  While the unit cell of such an arrangement is much smaller than a wavelength, the entire configuration requires many elements in series and can result in a very long structure relative to the wavelength.

In this work we have employed hydrophobic silica aerogel which has a static density of 107 kg/m$^{3}$ and an (optical) refractive index of 1.03. This silica aerogel has a high resistance for water and moisture due to its hydrophobicity, a feature allowing the disks to be fabricated using water jet cutting techniques.  Their properties are stable in any climate, a feature contributing the high reliability for the analysis provided below.

%%%%%%%%%%%%%%%%%%%%%%%%%%%%%%%%%%%%%%%%%%%%%%%%%%%%%%%%%%%%%%
\section{Theoretical formulation} \label{Sec:Theory}

In many acoustic metamaterial configurations, including LRAMs and acoustic transmission line arrangements, thin elastic disks and membranes are used as sub-wavelength stiffness elements.  This ubiquitous implementation arises from the fact that the time-harmonic displacement of an elastic disk is proportional to the flexural stiffness in the quasi-static (low frequency) limit, and therefore yields an analogous inductive behavior equivalent to that of a mechanical spring.  In general, however, the specific acoustic impedance of a lossless elastic plate can be expressed in terms of the mass per area $M_{\mathrm{plate}}$ and compliance $C_{\mathrm{plate}}$ as\cite{Blackstock}
\begin{align} 
Z_{\mathrm{plate}} &= j\omega M_{\mathrm{plate}} + \frac{1}{j\omega C_{\mathrm{plate}}} \notag \\*
&=  j\omega M_{\mathrm{plate}} \left[1 - \left(\frac{\omega_{\mathrm{res}}}{\omega}\right)^{2} \right] \equiv  j\omega \, M_{\mathrm{eff}}(\omega), \label{Eq:Zkm}
\end{align}
\noindent where $\omega_{\mathrm{res}} \!=\! (M_{\mathrm{plate}} C_{\mathrm{plate}})^{-1/2}$ is the angular resonance frequency of the plate.  Written in this form, it is apparent that the stiffness-controlled response of the elastic plate can equivalently be treated as a frequency-dependent effective mass, $M_{\mathrm{eff}}$, which is negative for $\omega \!<\!\omega_{\mathrm{res}}$.  

For canonical shapes and idealized boundary conditions, analytic expressions have been developed to describe the flexural wave motion of elastic solids.  For the flexural motion of thick plates, effects from shear deformation and rotational inertia become important, which can be formulated using Mindlin theory\cite{Graff}.  For flexural waves in a thick circular elastic disk, the modal displacement, $w_{n}$ of the plate is given by\cite{Lee2005,Hettema1988}
\begin{align} 
	w_{n}(r,\theta) &= \left[ A_{1} J_{n}(\delta_{1}\frac{r}{a}) + A_{2}I_{n}(\delta_{2}\frac{r}{a}) \right] \cos(n\theta), \label{Eq:PlateDisp} \\
 	\delta_{1}^{2} &= \frac{1}{2}\lambda^{4} \left[ \left(R + S\right) + \sqrt{(R - S)^{2} +4\lambda^{-4}} \right], \label{Eq:Delta1} \\
	\delta_{2}^{2} &= \frac{1}{2}\lambda^{4} \left[  \sqrt{(R - S)^{2} +4\lambda^{-4}} - \left(R + S\right) \right], \label{Eq:Delta2} \\
	R &= \frac{1}{12}\left(\frac{h}{a} \right)^{2}, \label{Eq:MindlinR} \\
	S &= \frac{D}{\mu h (\kappa a)^{2}} = \frac{1}{6(1-\nu)\kappa^{2}} \left(\frac{h}{a} \right)^{2}, \label{Eq:MindlinS} \\
	\lambda^{4} &= \frac{\rho h a^{4} \omega^{2}}{D}, \label{Eq:MindlinLambda} \\
	D &= \frac{E h^{3}}{12(1 - \nu^{2})}, \label{Eq:MindlinD}
\end{align}
\noindent where $J_{n}$ and $I_{n}$ denote the Bessel function and modified Bessel function of the first kind, $a$ is the plate radius, $h$ is the plate thickness, $\rho$ is the mass density, $E$ is Young's modulus, $\mu$ is the shear modulus, $\nu$ is Poisson's ratio, $\omega$ is the angular frequency and $\kappa \!=\! \pi/\sqrt{12}$ is the shear correction factor.  Note that for axisymmetric loading, such as that encountered in an acoustic impedance tube, the only non-zero mode is $n \!=\! 0$.  

Due to the clamped boundary conditions corresponding to $w(a) \!=\! w^{\prime}(a) \!=\! 0$, the characteristic equation for the flexural resonance frequencies is given by
\begin{equation} \label{Eq:ClampedCharEq}
	J_{1}(\delta_{1}) I_{0}(\delta_{2}) + J_{0}(\delta_{1}) I_{1}(\delta_{2}) = 0.
\end{equation}
\noindent As observed from Eq.~(\ref{Eq:Zkm}), the resonance frequency denotes a critical point in the effective acoustic properties, and even though Eq.~(\ref{Eq:ClampedCharEq}) provides the exact solution, it must be solved numerically. To better understand the relationship between the material properties and dimensions of the plate and the resonance frequencies, an approximate analytic expression is sought.  To proceed, it will be assumed that the flexural wavenumbers $\delta_{1}$ and $\delta_{2}$ are sufficiently high that the large argument approximations can be used for the Bessel and modified Bessel functions can be used, namely, \cite{Abramowitz}
\begin{equation} \label{Eq:BesslFunctApprox}
	\frac{J_{1}(x)}{J_{0}(x)} \approx \tan(x - \frac{\pi}{4}), \quad \frac{I_{1}(x)}{I_{0}(x)} \approx 1.
\end{equation}

Even in the case of large wavelengths relative to the thickness of the plate (and therefore in the low frequency limit in terms of the acoustic waves in the surrounding fluid), the flexural modes occur at frequencies where the wavelengths are on the order of the diameter of the disk or less, suggesting that the assumption above is reasonable for the case of flexural acoustic metamaterial elements being investigated in this work.  Note that because the ratio of the modified Bessel functions given in Eq.~(\ref{Eq:BesslFunctApprox}) is approximately equal to unity, the resonance frequency under these conditions only depends on $\delta_{1}$.  Application of the approximate expressions in Eq.~(\ref{Eq:BesslFunctApprox}) to Eq.~(\ref{Eq:ClampedCharEq}) yields
\begin{equation} \label{Eq:ClampedCharEqApprox}
	\tan(\delta_{1}) \approx 0,
\end{equation}
\noindent for which $\delta_{1} \!=\! m\pi$ with $m$ being a non-zero integer.  Equation~(\ref{Eq:Delta1}) can therefore be written as
\begin{equation} \label{Eq:Delta1Approx}
	\frac{1}{2}\lambda^{4} \left[ \left(R + S\right) + 2\lambda^{-2}\sqrt{1 + \Delta} \right] = (m\pi)^{2},
\end{equation}
\noindent where
\begin{equation} \label{Eq:Delta1ApproxDelta}
	\Delta = \frac{1}{4} \lambda^{4} (R - S)^{2} = \frac{3}{144} (k_{\mathrm{p}} h)^{2} \left[ \frac{2}{(1 - \nu)\kappa^{2}} - 1\right]^{2},
\end{equation}
with $k_{\mathrm{p}} \!=\! \omega/c_{\mathrm{p}}$ denoting the \emph{compressional} plate wave number with plate wave speed,
\begin{equation} \label{Eq:PlateSpeed}
	c_{\mathrm{p}} = \sqrt{ \frac{E}{\rho (1 - \nu^{2})} }.
\end{equation}
\noindent Typical values of compressional plate wave speeds are usually orders of magnitude higher than flexural wave speeds, and thus the corresponding wavenumbers are much lower.  As a result, one expects that $k_{\mathrm{p}}h \ll 1$ and likewise $\Delta \ll 1$, in which case Eq.~(\ref{Eq:Delta1Approx}) can be simplified to give an expression for the flexural resonance frequency of the $m^{\mathrm{th}}$ mode,
\begin{equation} \label{Eq:FresApprox}
	f_{\mathrm{res}}^{(m)} \!=\! \frac{1}{4 \pi \sqrt{3}} \frac{ c_{\mathrm{p}} h }{a^{2}} \frac{1}{R \!+\! S} \left[ \sqrt{1 + 2(m\pi)^{2}(R \!+\! S)} \!-\! 1 \right].
\end{equation}

\begin{figure}[t!]
	\includegraphics[width=0.99\columnwidth, height=0.7\textheight, keepaspectratio]{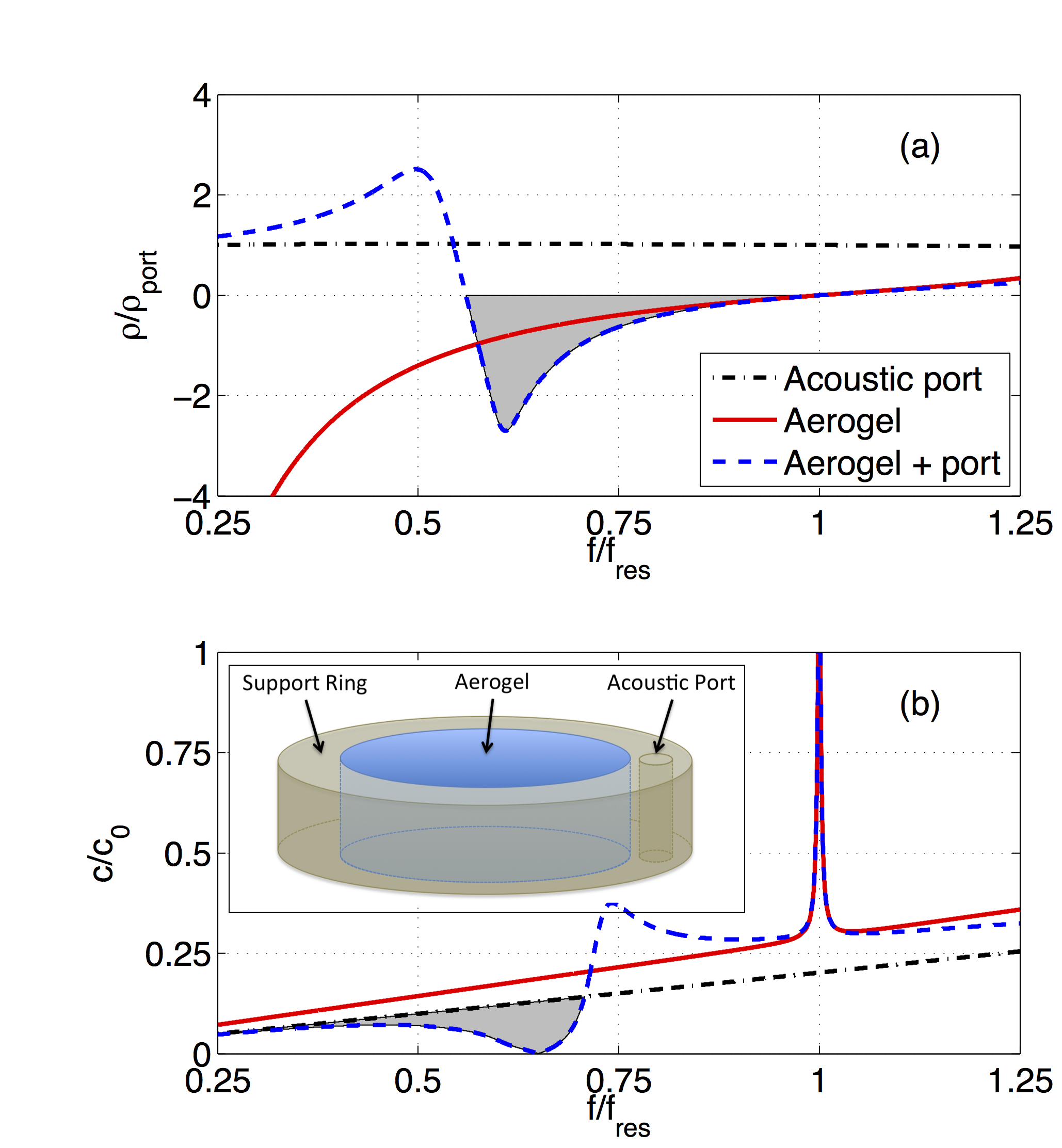}
	\caption{ (Color online) Normalized plots versus frequency for an aerogel disk (solid red line), an acoustic port with a rigid baffle (dash-dotted black line) and an aerogel disk combined in parallel with an acoustic port (dashed line) for (a) effective density relative to that of a baffled acoustic port, and (b) effective sound speed relative to that of the surrounding fluid medium.  The shaded regions denote broadband regions of extreme effective properties, namely (a) negative density, and (b) non-resonant slow sound propagation.  An illustration of the parallel arrangement of an aerogel disk with an acoustic port is shown in the inset of (b).}
	\label{Fig:Config}
\end{figure}

In addition to the flexural resonance frequency, the particular solution to the displacement is necessary to calculate the effective acoustic properties of the flexural disk.  Assuming a time harmonic pressure $P$ applied across the face of the disk and assuming clamped edges, the displacement becomes
\begin{equation} \label{Eq:DispClamped}
	w(r) \!=\! \frac{P}{\rho h \omega^{2}} \!\! \left[ \frac{ I_{1}(\delta_{2}) J_{0}(\delta_{1} \frac{r}{a})  + J_{1}(\delta_{1}) I_{0}(\delta_{2} \frac{r}{a}) }{ I_{1}(\delta_{2}) J_{0}(\delta_{1}) + J_{1}(\delta_{1}) I_{0}(\delta_{2})  } - 1 \right].
\end{equation}
\noindent Although Eq.~(\ref{Eq:DispClamped}) gives an expression for the displacement at any given radius $r$, it is actually the ensemble of the displacement over all the points on the surface which will be measured via the reflected or transmitted acoustic waves at some distance from the disk.  Therefore, a more useful quantity is the spatial average of the displacement, which can be obtained from Eq.~(\ref{Eq:DispClamped}) such that
\begin{equation} \label{Eq:AvgDispClamped}
	w_{\mathrm{avg}} \!=\! \frac{P}{\rho h \omega^{2}} \!\! \left[ \frac{ 2(\frac{1}{\delta_{1}} \!+\! \frac{1}{\delta_{2}})J_{1}(\delta_{1}) I_{1}(\delta_{2}) }{ I_{1}(\delta_{2}) J_{0}(\delta_{1}) + J_{1}(\delta_{1}) I_{0}(\delta_{2})  } - 1 \right],
\end{equation}
\noindent from which one can obtain the average acoustic impedance for a thick clamped circular plate
\begin{equation} \label{Eq:Zplate}
	Z_{\mathrm{plate}} \!=\! \frac{j\omega M_{\mathrm{plate}}}{ 1 \!-\! 2(\frac{1}{\delta_{1}} \!\!+\!\! \frac{1}{\delta_{2}}) \!\!\left[ \frac{ J_{1}(\delta_{1}) I_{1}(\delta_{2}) }{ I_{1}(\delta_{2}) J_{0}(\delta_{1}) + J_{1}(\delta_{1}) I_{0}(\delta_{2})  } \right]} \!=\!  j\omega M_{\mathrm{eff}},
\end{equation}
\noindent where $M_{\mathrm{plate}} \!=\! \rho h$ is the acoustic mass of the plate.  Although not as obvious as the form presented in Eq.~(\ref{Eq:Zkm}), the expression presented in Eq.~(\ref{Eq:Zplate}) also yields a negative effective mass below the first flexural resonance of the elastic plate, which is illustrated by the red solid line in Fig.~\ref{Fig:Config}(a) for the case of a circular silica aerogel disk.  The resulting negative effective density extends over the entire range below resonance and approaches $-\infty$ as the frequency goes to zero.

%%%%%%%%%%%%%%%%%%%
%%% Figure 1
%%%%%%%%%%%%%%%%%%%

Such a highly dispersive and divergent behavior is not ideal, particularly for broadband applications or if any type of acoustic impedance matching is desired.  However, this effective mass can be readily modified by placing a positive acoustic mass (such as an acoustic port consisting of an air-filled hole in the support ring) in parallel with the negative dynamic mass of the plate.  The resulting frequency dependence on the effective mass density is presented in Fig.~\ref{Fig:Config}(a), in which a circular disk (red solid line) is combined in parallel with an acoustic port (black dash-dotted line) to obtain the effective density denoted by the dashed line.  This parallel configuration of the aerogel disk and acoustic port is illustrated in the inset of Fig.~\ref{Fig:Config}(b).  Arranged in such a manner, the acoustic port will short circuit the plate as it approaches extremely large values, and allow for the magnitude and bandwidth of the negative dynamic mass (denoted by the shaded region) to be controlled using the same plate and only varying the size and number of ports.

A similar trend is observed for the effective sound speed, which is illustrated in Fig.~\ref{Fig:Config}(b).  As in Fig.~\ref{Fig:Config}(a), the results for a circular elastic disk (red solid line) is arranged in parallel with an acoustic port (black dash-dotted line) to obtain the effective density denoted by the dashed line.  The effective sound speed of the combined parallel arrangement follows that of the acoustic port at low frequencies, slowly decreasing towards zero before increasing again, producing a broad non-resonant region where slow sound propagation occurs (denoted by the shaded region), slower than even that expected for the case of a acoustic port.  As the frequency increases, the effective sound speed becomes dominated by the elastic disk in the vicinity of the flexural resonance of the elastic plate, which occurs at $f \!=\! f_{\mathrm{res}}$.  Near this resonance, a large increase in the effective sound speed is observed as the effective density passes through zero, with the peak value limited by the losses in the system.

%%%%%%%%%%%%%%%%%%%
%%% Figure 2
\begin{figure*}[t!]
	\includegraphics[width=0.99\textwidth, height=0.7\textheight, keepaspectratio]{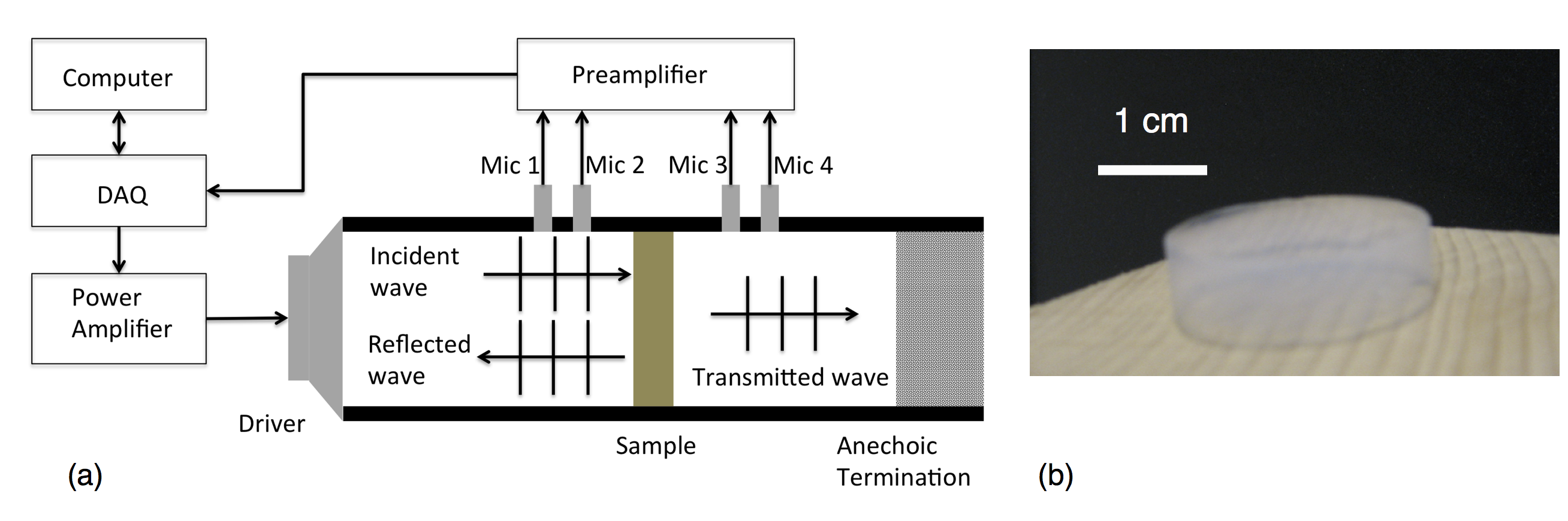}
	\caption{(Color online) (a) Diagram of the experimental setup using an air-filled acoustic impedance tube, and (b) a photo of a silica aerogel sample examined in this work. } 
	\label{Fig:Setup}
\end{figure*}
%%%%%%%%%%%%%%%%%%%

%%%%%%%%%%%%%%%%%%%%%%%%%%%%%%%%%%%%%%%%%%%%%%%%%%%%%%%%%%%%%%
\section{Experimental setup} \label{Sec:Setup}
Previous approaches for the acoustic characterization of silica aerogels have been based on ultrasonic time-of-flight measurements, which have produced experimental data for the compressional and shear properties of silica aerogels \cite{Gross1988,Gomez2002,AerogelHandbookChp22,Xie1998}.  Unfortunately, such time-of-flight measurements do not allow for one to examine the silica aerogel as part of the metamaterial structure, requiring the use of acoustic waves with wavelengths which are much smaller than the thickness of the sample for accurate time-of-flight characterization.  Although most of these previous investigations involved compressional and shear waves, Rayleigh surface waves have also been examined, which were observed to be less than 50 m/s, significantly lower than the compressional wave speed\cite{Rogers1999}.  Likewise, as can be seen from Fig.~\ref{Fig:Config}, the effective properties due to the flexural behavior of the aerogel disks at large wavelengths (low frequencies) relative to the size of the disk are dramatically different than those of the static density and compressional wave speed obtained via time-of-flight measurements for silica aerogel \cite{Gross1988}.

Alternatively, an experimental setup was needed that could examine the effective acoustic behavior of the ensemble arrangement including the silica aerogel disk, at frequencies which were sufficiently low to be within the homogenization limit for use in the microstructure of an acoustic metamaterial.  The experimental investigation of the silica aerogel samples in this work was performed using an air-filled acoustic impedance tube.  The experimental setup is illustrated in Fig.~\ref{Fig:Setup}(a), which shows a standard 4-microphone configuration \cite{Salissou2010} for the measurement of acoustic properties of a given sample.  The acoustic properties, namely the complex-valued acoustic impedance and wavenumber, were obtained as a function of input frequency from spectral measurements of the magnitude and phase of both the reflected and transmitted acoustic pressure obtained from the 4 microphones.  

While this particular type of acoustic apparatus has been utilized for many decades, such work has traditionally focused on simply measuring transmission loss through an absorptive sample \cite{Song2000}.  In the last several years, this technique has been expanded to acoustic metamaterials, with a particular emphasis on extraction of the effective complex acoustic properties of the acoustic metamaterial sample \cite{Fokin2007, Guild2014, Guild2015}.  The method for the extraction of these properties can be found in previous works \cite{Guild2014}, based on the complex reflection and transmission pressure coefficients, which are given by
\begin{equation} \label{Eq:Rtot}
	R_{\mathrm{tot}} = \frac{1}{2}\frac{ Z_{\mathrm{tot}}}{ Z_{0} } \! \left[ 1 + \frac{1}{2}\frac{ Z_{\mathrm{tot}} }{Z_{0}} \right]^{-1}.
\end{equation}
\begin{equation} \label{Eq:Ti}
	T_{\mathrm{tot}} = \left[ 1 + \frac{j \omega}{2 Z_{0}} \Big( M_{\mathrm{eff}} \!+\! M_{\mathrm{port}} \Big) \right]^{-1}, 
\end{equation}
\noindent where $Z_{\mathrm{tot}}$ is the total acoustical impedance seen at the face of the sample, $Z_{0}$ is the acoustical impedance of air, and $M_{\mathrm{eff}}$ and $M_{\mathrm{port}}$ are the effective acoustic mass of the aerogel disk and port, respectively.

This inverse process of using spectral acoustic measurements to obtain the complex-valued acoustic properties has presented significant challenges, including ambiguity in identifying unique solutions \cite{Fokin2007} and a high sensitivity in the extracted properties to even low levels of noise in the spectral data obtained from the microphones \cite{Guild2014}.  Although these challenges have recently been overcome for samples such as sonic crystals\cite{Guild2014, Guild2015} with a relatively low acoustic impedance relative to the surrounding air, acoustic metamaterials consisting of solid elastic structures, even relatively soft ones such as silica aerogels, with acoustic impedances hundreds of times larger than that of air present unique challenges using this technique.

As a result of this high acoustical input impedance from the sample, the small but finite leakage from the impedance tube becomes a significant source of error in the measurement and must be accounted for.  Typically, this leakage has been observed as a result of improper sealing and mounting of the sample, with leakage of acoustic energy from the reflected side of the sample to the transmitted side.  However, in the case of rigidly mounted high-impedance samples, this leakage is primarily due to the microphones, which consist of thin diaphragms and small acoustic ports for pressure equalization, with the main source of this leakage occurring on the reflected side of the sample.  As a result, two distinct differences arise for this case of microphone pressure leakage compared with that due to improper mounting: (1) the leakage occurs in the reflected pressure measurements only, with negligible effects on the transmitted side, and (2) for large impedances, the leakage should be independent of the specific sample or its particular mounting in the impedance tube.  

%%%%%%%%%%%%%%%%%%%
%%% Figure 3
\begin{figure*}[t!]
	\includegraphics[width=0.99\textwidth, height=0.7\textheight, keepaspectratio]{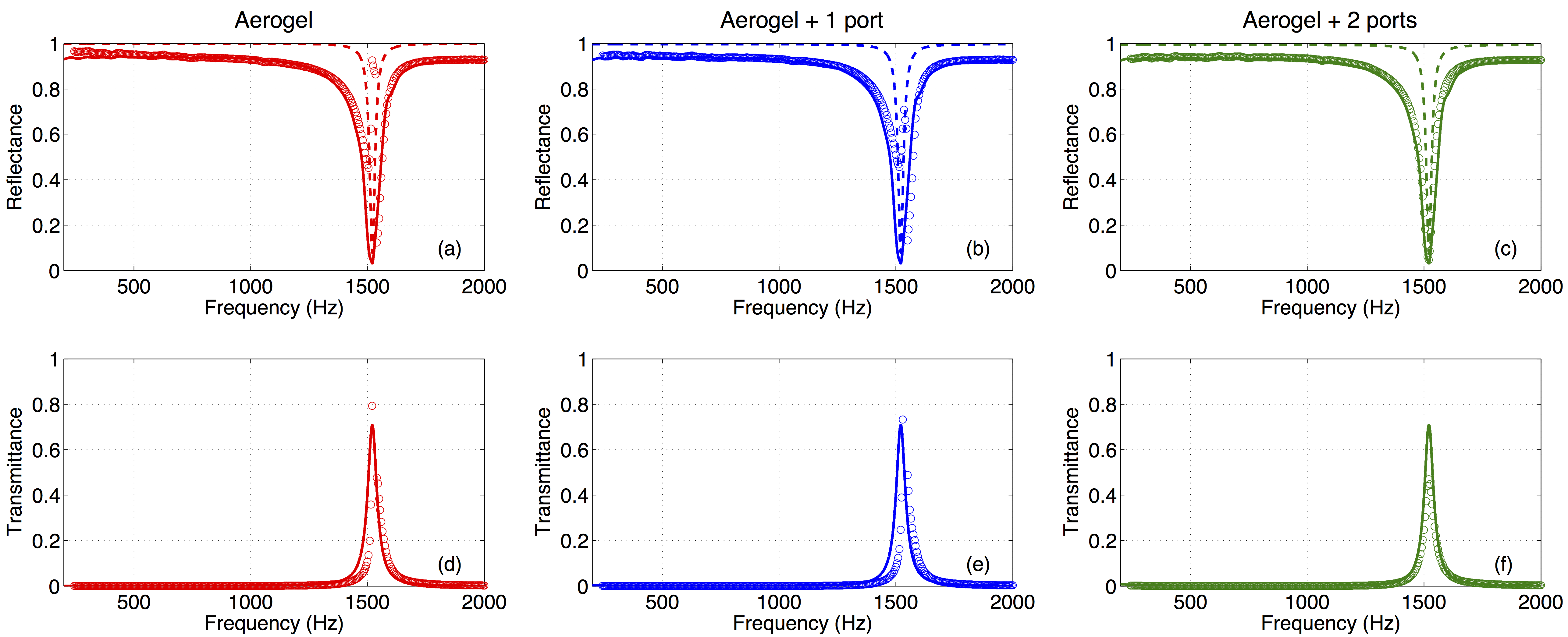}
	\caption{ (Color online) Comparison of experimental data with the theoretical model presented in Sec.~\ref{Sec:Theory} for the acoustic reflectance (a)-(c), and transmittance (d)-(f).  Experimental data is denoted by circles, and the theoretical results are presented with and without corrections due to the pressure leakage through the microphones, denoted by the solid and dashed lines, respectively. } 
	\label{Fig:DataRT}
\end{figure*}
%%%%%%%%%%%%%%%%%%%

Under these conditions, the total acoustical impedance seen at the front of the sample can be written as 
\begin{equation} \label{Eq:ZtotLeak}
	Z_{\mathrm{tot}} = \left[ \frac{\phi_{\mathrm{plate}}}{Z_{\mathrm{plate}}} \!+\! \frac{\phi_{\mathrm{port}}}{Z_{\mathrm{port}}} \!+\! \frac{\phi_{\mathrm{leak}}}{Z_{\mathrm{leak}}} \right]^{-1},
\end{equation}
\noindent where $\phi$ denotes the filling fraction of each component, $Z_{\mathrm{port}}$ is the impedance of the acoustic port including thermosviscous effects \cite{Kinsler}, and $Z_{\mathrm{leak}}$ is the pressure leakage impedance.  
\par
To determine the leakage impedance, one can consider the measurement of a rigid or nearly-rigid sample without any acoustic ports, in which case the total impedance is simply $Z_{\mathrm{tot}} \!=\! Z_{\mathrm{leak}}/\phi_{\mathrm{leak}}$.  In the absence of the pressure leakage, all the sound should be reflected and the reflectance should be unity; however, in the presence of the leakage, the measured reflection coefficient for a rigid or nearly-rigid sample, $R_{\mathrm{meas}}$, will exhibit a reduction in the magnitude (and a phase difference as well).  An expression for the leakage effects can be obtained using Eq.~(\ref{Eq:Rtot}) in terms of the measured complex reflection coefficient $R_{\mathrm{meas}}$ as
\begin{equation} \label{Eq:Zleak}
	\frac{Z_{\mathrm{leak}}}{\phi_{\mathrm{leak}}} = \frac{2 Z_{0} R_{\mathrm{meas}}}{1 - R_{\mathrm{meas}}}.
\end{equation}

Samples consisting of silica aerogel disks measuring 2.43 cm in diameter and 1.1 cm thick were tested in a 3.5 cm diameter air-filled acoustic impedance tube, as pictured in Fig.~\ref{Fig:Setup}(b).  The acoustic impedance tube used in this work consists of a circular tube having an inner diameter of 3.5 cm.  To mount the silica aerogel disks in the impedance tube, wooden rings were machined to support the aerogel disks and provide an acoustic baffle.  Due to the brittleness of the aerogel disks, the holes were drilled into the wooden ring to create the acoustic ports, each with a diameter of 1 mm.  To obtain the acoustic characterization of the aerogel samples, broadband noise is generated using an electromechanical driver at one end of the tube, and measured using 0.50 inch (1.27 cm) diameter G.R.A.S. condenser microphones.  The microphones are arranged in a standard 4-microphone configuration\cite{Salissou2010}, allowing for the reflection and transmission pressure coefficients to be directly determined using a transfer-matrix technique\cite{Song2000}.  From this set of measurements, the complex impedance and wavenumber were obtained for the range 300--2000 Hz, in a similar manner to previous experimental work on acoustic metamaterial samples\cite{Guild2014,Guild2015}.

%%%%%%%%%%%%%%%%%%%%%%%%%%%%%%%%%%%%%%%%%%%%%%%%%%%%%%%%%%%%%%
\section{Results} \label{Sec:Results and discussion}

The experimental data for the reflectance are shown in Fig.~\ref{Fig:DataRT}(a)-(c), with the corresponding transmittance data shown in Fig.~\ref{Fig:DataRT}(d)-(f), for the nominal aerogel sample plus with 1 and 2 acoustic ports, respectively.  
These measured values are relatively constant with frequency except in the vicinity of 1500 Hz, where there is a rapid decrease in reflectance with a corresponding increase in the transmittance, due to the flexural resonance of the circular aerogel disk.  This anomalous increase in the transmission of acoustic energy in the experimental data far exceeds that expected based on quasi-static homogenization theory for such a large impedance contrast with the surrounding air, and corresponds to a region of  extraordinary transmission.

%%%%%%%%%%%%%%%%%%%
%%% Table 1
\begin{table}[b!]
	\begin{ruledtabular}
		\begin{tabular}{ccc}
			\bf{$f_{\mathrm{res}}$ (Hz)} & \bf{$E$ (MPa)} & \bf{$\rho$ ($kg/m^{3}$)} \\ \hline
			$1420$ & $0.569$ & $107$ 
		\end{tabular}
	\end{ruledtabular}
	\caption{Measured values for the flexural resonance frequency, Young's modulus $E$ and density $\rho$ for the silica aerogel examined in this work.  The loss factor of the aerogel was observed to be $0.005$ based on the reflectance and transmittance data presented in Fig.~\ref{Fig:DataRT}.  The Poisson's ratio was estimated to be $0.21$ based on Gross \emph{et al.}\cite{Gross1988}.}
	\label{Tab:Props}
\end{table}
%%%%%%%%%%%%%%%%%%%

Based on the measured data, the flexural resonance frequency is obtained, and the Young's modulus of the aerogel sample can be obtained based on the theoretical expression for the resonance frequency given by Eq.~(\ref{Eq:FresApprox}).  The measured properties of the aerogel sample examined in this work are tabulated in Table~\ref{Tab:Props}.  While these measured values are somewhat lower than some other similar aerogel samples\cite{Gross1988}, the results reported here fall within the accepted range of measured aerogel properties\cite{Gronauer1986}.

\begin{figure*}[t!]
	\includegraphics[width=0.99\textwidth, height=0.7\textheight, keepaspectratio]{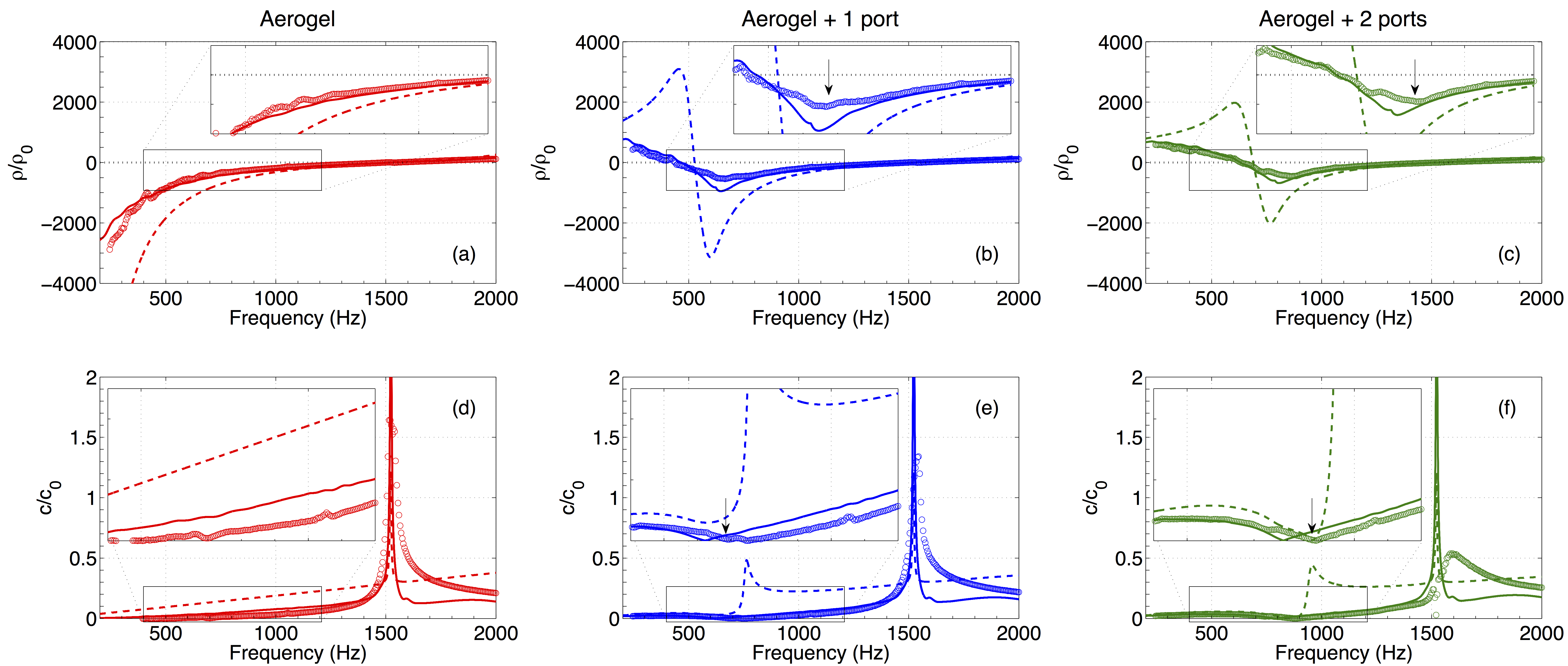}
	\caption{ (Color online) Comparison of experimental data with the theoretical model presented in Sec.~\ref{Sec:Theory} normalized relative to the properties of air ($\rho_{0} \!=\! 1.21$ kg/m$^{3}$, $c_{0} \!=\! 343$ m/s), for the effective density (a)-(c), and effective sound speed (d)-(f).  Experimental data is denoted by circles, and the theoretical results are presented with and without corrections due to the pressure leakage through the microphones, denoted by the solid and dashed lines, respectively. The arrows denote extreme values, namely the peak negative density in (b) and (c), and near-zero slow sound propagation in (e) and (f).} 
	\label{Fig:DataRhoC}
\end{figure*}

From these measured values, an elastic model based on the theoretical formulation presented in Sec.~\ref{Sec:Theory} was calculated and compared to the data.  Modeled results obtained without accounting for the pressure leakage effects are denoted by the dashed line in Fig.~\ref{Fig:DataRT} and Fig.~\ref{Fig:DataRhoC}.  While there is excellent agreement with the transmittance shown in Fig.~\ref{Fig:DataRT}(d)-(f), the theoretical reflectance shown in Fig.~\ref{Fig:DataRT}(a)-(c) predicts unity away from the flexural resonance, a value which is not observed in the measurements.  While this difference between the modeled and measured values represents less than a 10\% error over most of  the frequency band under investigation, this leads to a significant variation in the resulting extracted effective acoustic properties, as illustrated in Fig.~\ref{Fig:DataRhoC}.  This is particularly the case below the flexural resonance frequency, for which the variation between the theoretical model without accounting for the leakage differs by up to an order of magnitude from that observed in the experimental data.

Accounting for the leakage through the use of Eqs.~(\ref{Eq:ZtotLeak}) and (\ref{Eq:Zleak}), this observed difference in the reflectance can be correctly modeled, as shown by the solid lines in Fig.~\ref{Fig:DataRT} and Fig.~\ref{Fig:DataRhoC}.  In addition to the improved agreement in the reflectance presented in Fig.~\ref{Fig:DataRT}(a)-(c), excellent agreement is also maintained with the transmittance in Fig.~\ref{Fig:DataRT}(d)-(f).  The theoretical model including pressure leakage and the measured data of the extracted effective mass density and sound speed are in excellent agreement, with the model capturing the correct magnitudes and frequency-dependence of these dynamic properties.  Even with the limitations of the experimental apparatus due to the finite acoustical impedance of the microphones and the resulting acoustic pressure leakage, one is still able to observe an extremely large dynamic range of extracted acoustic properties, on the order of thousands of times that of the ambient air.

In addition to the agreement between the experimental data and theoretical model, the results presented in Fig.~\ref{Fig:DataRhoC} provide valuable information regarding the wide range of effective properties which can be attained using the soft acoustic metamaterial arrangement illustrated in Fig.~\ref{Fig:Config}.  In particular, it can be observed that the overall trends as a function of frequency follow those predicted in Fig.~\ref{Fig:Config}(a) and (b) for silica aerogel disks with and without parallel arrangements with acoustic ports, for the effective mass density and sound speed, respectively.  In Fig.~\ref{Fig:DataRhoC}(a), the effective density for the data obtained for the silica aerogel disk is observed to be positive above the flexural resonance frequency  (around 1500 Hz), passing through zero, and then negative below the flexural resonance frequency and  tending towards $-\infty$ as frequency decreases.  Conversely, the effective sound speed shown in Fig.~\ref{Fig:DataRhoC}(d) increases significantly as the effective density passes through zero near the flexural resonance frequency, and decreases towards zero as the frequency decreases.  These effective properties, which arise from use of the flexural motion of the silica aerogel as a ``hidden degree of freedom"\cite{Milton2007} in the otherwise 1D planar arrangement of the acoustic impedance tube, lead to these extreme effective properties which different greatly from the static properties of silica aerogel.

%%%%%%%%%%%%%%%%%%%
%%% Figure 4
%%%%%%%%%%%%%%%%%%%

Likewise, the combination of the negative effective density of the silica aerogel disk with the positive effective density of one (or more) acoustic ports yields much more uniform and less dispersive regions of negative density, which are illustrated in Fig.~\ref{Fig:DataRhoC}(b) and (c) for the case of 1 and 2 acoustic ports, respectively.  As described above, the parallel arrangement of the silica aerogel disk and the acoustic ports leads to the effective properties being dominated by the acoustic port at lower frequencies, leading to this change in the effective density as a function of frequency over this range below the flexural resonance frequency.  In the vicinity of the flexural resonance, however, the effective properties are dominated by the silica aerogel disk, and therefore the same density-near-zero region is observed.  

Similarly, the effective sound speed is presented in Fig.~\ref{Fig:DataRhoC}(e) and (f) for the case of silica aerogel with 1 and 2 acoustic ports, respectively.  In the vicinity of the flexural resonance, a similar spike in the effective sound speed is observed (corresponding to the effective density passing through zero) as was observed for the silica aerogel disk without any acoustic ports.  In addition to this, a dip in the effective sound speed is observed well below the flexural resonance, leading to a region of zero and near-zero effective wave speed in the acoustic metamaterial structure.  Previous experimental investigations of slow sound have been observed over relatively narrow bands due to the resonant physical mechanisms employed, whereas the results presented in this work appear to be the first to experimentally demonstrate a broad region of non-resonant slow sound propagation.

Similar trends in the data described above were observed for the silica aerogel disk combined with either 1 or 2 acoustic ports.  One interesting point of distinction that can be noted is that the addition of more acoustic ports (and thus a larger, positive effective acoustic mass) results in an \emph{increase} in the frequency at which the minimum of the negative density and region of slow sound occur.  This stands in contrast with traditional resonant acoustic metamaterials, such as those utilizing simple harmonic oscillators, for which the addition of mass tends to \emph{decrease} the resonant frequency and the corresponding frequencies of negative effective density.  This difference for the soft acoustic metamaterial arrangement investigated here arises due to the parallel arrangement of the different components, compared with the tradition series arrangement of mass-spring and transmission line acoustic metamaterials.  Although the soft acoustic metamaterials examined in this work were limited to relatively simple lumped elements and canonical geometries, this principle can be extended to a wide range of more elegant acoustic elements allowing for a vast range of tunable exotic properties in a compact, conformal design.

%%%%%%%%%%%%%%%%%%%%%%%%%%%%%%%%%%%%%%%%%%%%%%%%%%%%%%%%%%%%%%
\section{Conclusion} \label{Sec:Conclusion}
In conclusion, silica aerogel disks have been examined theoretically and experimentally as building units of a soft acoustic metamaterial.  It has been shown that the combination of the flexural motion of the silica aerogel combined with the acoustic mass of one or more ports leads to a configuration with broadband negative dynamic density, density-near-zero regions and non-resonant broadband slow sound propagation.  The use of silica aerogel as part of a soft acoustic metamaterial structure with subwavelength thickness was examined theoretically and experimentally.  Significant challenges were overcome to obtain direct measurements of the effective density and sound speed for such high impedance metamaterial elements, which was achieved using an air-filled acoustic impedance tube and correcting for the inherent pressure leakage from the microphones.  The experimental measurements were found to be in very good agreement with the expected theoretical results.   Unlike acoustic metamaterials utilizing mass-spring resonators, the soft acoustic metamaterials described in this work move beyond this framework by creating the dynamics from the flexural motion of a soft elastic structure, while offering the tunability to achieve a wide range of desirable exotic properties with a single subwavelength element.

\section*{Acknowledgements}
This work was supported by the U.S. Office of Naval Research and by the Spanish \emph{Ministerio de Econom\'{i}a y Competitividad} (MINECO) under grant number TEC2014-53088. The authors wish to acknowledge Encarna G. Villora and Kiyoshi Shimamura for their help in aerogel fabrication and handling.

%%%%%%%%%%%%%%%%%%%%%%%%%%%%%%%%%%%%%%%%%

%\bibliography{AerogelPaperBib}

%\newpage
%\printtables
%\newpage
%\printfigures
%\newpage

\end{document}